\documentclass[aps,prx,reprint,floatfix,superscriptaddress,amsmath,amssymb,longbibliography]{revtex4-2}

\usepackage{graphicx}
\usepackage{amsthm}
\usepackage{bm}
\usepackage{xcolor}
\definecolor{linkblue}{HTML}{1A4F8B}
\usepackage[colorlinks=true,linkcolor=linkblue,citecolor=linkblue,urlcolor=linkblue]{hyperref}
\hypersetup{pdftitle={What a collective can hold in common: a gauge framework for private representations},pdfauthor={Mohammad Salahshour},pdfsubject={Relational geometry of collective compatibility and feature recovery},pdfkeywords={collective behavior, private representations, gauge covariance, holonomy, connection Laplacian}}

\newtheorem{theorem}{Theorem}
\newtheorem{proposition}{Proposition}
\newtheorem{corollary}{Corollary}
\theoremstyle{remark}
\newtheorem{remark}{Remark}

\newcommand{\Fix}{\operatorname{Fix}}
\newcommand{\Hol}{\operatorname{Hol}}
\newcommand{\tr}{\operatorname{tr}}
\graphicspath{{./}}
\newcommand{\SIref}[1]{\csname SI@#1\endcsname}
\expandafter\def\csname SI@eq:s_conjugation_distance\endcsname{S10}
\expandafter\def\csname SI@fig:s1\endcsname{S1}
\expandafter\def\csname SI@fig:s10\endcsname{S11}
\expandafter\def\csname SI@fig:s2\endcsname{S2}
\expandafter\def\csname SI@fig:s3\endcsname{S3}
\expandafter\def\csname SI@fig:s7\endcsname{S7}
\expandafter\def\csname SI@fig:s8\endcsname{S9}
\expandafter\def\csname SI@fig:s9\endcsname{S10}
\expandafter\def\csname SI@fig:source_weighting\endcsname{S8}
\expandafter\def\csname SI@sec:s2\endcsname{S2}
\expandafter\def\csname SI@sec:s3\endcsname{S3}
\expandafter\def\csname SI@sec:s4\endcsname{S4}
\expandafter\def\csname SI@sec:s5\endcsname{S5}
\expandafter\def\csname SI@sec:s6\endcsname{S6}
\expandafter\def\csname SI@sec:s7\endcsname{S7}
\expandafter\def\csname SI@sec:s8\endcsname{S8}
\expandafter\def\csname SI@sec:s9\endcsname{S9}
\expandafter\def\csname SI@sec:s_categorical_features\endcsname{S5.3}
\expandafter\def\csname SI@sec:s_circle_features\endcsname{S5.2}
\expandafter\def\csname SI@sec:s_route\endcsname{S10}
\expandafter\def\csname SI@sec:s_spherical_features\endcsname{S7.3}

\begin{document}
\widowpenalty=3000
\clubpenalty=3000

\title{\texorpdfstring{What a collective can hold in common:\\
a gauge framework for private representations}{What a collective can hold in common: a gauge framework for private representations}}

\author{Mohammad Salahshour}
\email{salahshour.mohammad@gmail.com}
\affiliation{Department of Collective Behaviour, Max Planck Institute of Animal Behavior, 78464 Konstanz, Germany}
\affiliation{Centre for the Advanced Study of Collective Behaviour, University of Konstanz, 78464 Konstanz, Germany}
\affiliation{Department of Biology, University of Konstanz, 78464 Konstanz, Germany}

\date[]{}

\begin{abstract}
Many models of collective behavior write headings, beliefs, and meanings in
one experimenter-defined frame. Organisms do not live there: each represents
the world in a private space, and comparison requires translation. Consensus
becomes an existence problem before it becomes a dynamical one. Reciprocal
pairwise translations need not compose consistently around a loop. This
return transformation---the holonomy---can expose a mismatch no isolated
reciprocal pair can carry.
We develop a gauge-covariant theory in which physical predictions are
invariant under private relabeling. For reciprocal unitary translations,
the kernel of the connection Laplacian is isomorphic to the joint fixed
space of loop holonomies: a collective can hold in common exactly what all
its loops leave unchanged. A blind-subgroup criterion identifies invisible
defects. Common-frame comparison, such as in traditional models of collective
behaviour, occupies the flat sector. For independent uniform
translations from a finite group $G$, a connected graph with $E$ relations
among $N$ individuals is flat with probability $|G|^{-(E-N+1)}$. Exact cycle
spectra determine persistence under linear relaxation.
The consequence is operational as well as structural. For unrecorded routes
under a specified model with isotropic Gaussian source and readout noise,
we derive the smallest mean-squared error achievable by any decoder. At
fixed signal-to-noise ratio, the jointly preserved feature fraction fixes
its long-route limit. Networks with identical loop angles can differ in
recoverable content when their preserved axes differ. The relational
geometry of a collective sets both the states it can share and the content
it can recover.
\end{abstract}

\maketitle

\section{Introduction}
\label{sec:intro}

\begin{figure}[!t]
\centering\includegraphics[width=3.35in]{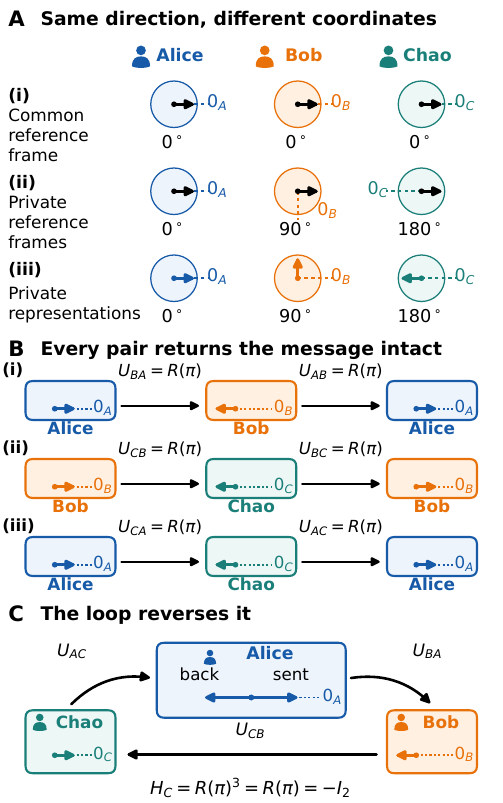}
\caption{\textbf{Every pair recovers the message; the triad reverses it.}
\textbf{A(i)--(ii)}~Black arrows show the food direction in a common frame;
dashed rays labeled $0$ mark angular zeros of private coordinates.
A common zero gives three $0^\circ$ coordinates; private zeros pointing
east, south, and west give representations $0^\circ,90^\circ,180^\circ$
(\textbf{A(iii)}).
\textbf{A(iii)}~Colored arrows show the same private representations with their zero
rays aligned: subjectively, they point east, north, and west, although all
represent objectively eastward food in the common frame. Coordinate conversions relate these descriptions to one shared
frame: every loop returns the identity, and the agreement is preserved.
\textbf{B(i)--(iii)}~Alice--Bob, Bob--Chao, and Alice--Chao each have
reciprocal unitary translations, chosen here as $U_{ij}=R(\pi)=-I_2$.
$U_{ij}$ maps $j$ to $i$; $R(\pi)$ is a planar half-turn.
Every pair understands in the round-trip sense: $U_{ji}U_{ij}=I_2$, so a
message returns intact, $x\mapsto-x\mapsto x$.
\textbf{C}~The same translations fail in the triad:
Alice $\to$ Bob $\to$ Chao $\to$ Alice gives
$x\mapsto-x\mapsto x\mapsto-x$, with $H_C=R(\pi)^3=-I_2$.
The message returns reversed in Alice's coordinates.
In \textbf{B--C}, colored arrows show private representations; black arrows show routing.
A shared directional state would require $a_A=-a_A$: no nonzero arrow
satisfies all three relations, although an unoriented axis survives.
Changing private zeros alone cannot create this loop mismatch.}
\label{fig:loop}
\end{figure}

A planet has no point of view. Nothing is lost by writing its state in
coordinates of our choosing, and for inanimate matter this is the first
lesson of physics. Models of collective behavior inherited the habit:
flocking models place every heading on one circle
\cite{Vicsek1995,TonerTu1995}, and consensus and opinion models place every
belief on one shared scale \cite{DeGroot1974,HegselmannKrause2002}.
Living beings are different from planets. An animal
carries a representation of the world built by its own sensing and history,
and that representation is not the modeler's lens but a variable the animal
itself holds \cite{SeeligJayaraman2015}. A model of collective behavior in which individuals carry a
viewpoint \cite{SalahshourCouzin2025,Sayin2025} is home to a different
physics from one in which they do not \cite{Vicsek1995,Couzin2002}: in the
first, collective behavior emerges when frame-carrying individuals interact,
with no additional rules of interaction \cite{SalahshourCouzin2025}; in the second, the rule was the
mechanism \cite{Vicsek1995,Couzin2002}. And once individuals are compared,
the shared frame does something further. It hands every pair a common
origin, and with it the guarantee that all of the group's comparisons fit
together. That guarantee is a physical assumption disguised as a convention
(Fig.~\ref{fig:loop}\textbf{A}).

These common-frame comparison rules, therefore, build compatibility in
\cite{DeGroot1974,HegselmannKrause2002}. It has thus been possible to ask how agreement is reached---how
individuals align, synchronize, or converge---without first asking whether
an agreement exists to be reached. The two questions have an order. Which
collective states are compatible with all of a group's relations comes
before how the group arrives at one of them. Under the shared frame the
prior question has a trivial answer, which is why it can go unasked; that is
not the same as its being settled. Here we ask what happens when the physics
is written in a frame that belongs to the individuals inside the system rather
than in a shared experimenter frame. Once comparison itself depends on
relations between private representations, agreement becomes an existence
problem before it becomes a dynamical one.

Physics has twice given up a privileged frame and asked instead what
survives every choice of one: for space and time \cite{Einstein1905}, and for
gauge fields \cite{Wilson1974,Kogut1979}, where the physical content is exactly
what no local convention can change. Here, we make the same move. It is available to
us for a simple reason: a living being has a worldview, so there is a
private convention to change in the first place. Each individual may
relabel its own representation however it likes; the dictionaries that
connect it to its neighbors change along with it; and nothing measurable
changes at all. In symbols, $a_i\mapsto h_ia_i$ and
$U_{ij}\mapsto h_iU_{ij}h_j^{-1}$, which is the substitution that defines a
gauge transformation on a lattice \cite{Wilson1974,Kogut1979}. Physical statements must therefore be
built from what this freedom cannot alter. This is a gauge-covariance
principle \cite{Wilson1974,Kogut1979}. It is not a metaphor imported from field theory: it is the same
requirement, and the objects that meet it are the same objects---parallel
transport \cite{SingerWu2012,Gao2021}, the composition of translations along a path;
holonomy \cite{Gao2021}, the return map around a loop; and the connection
Laplacian \cite{SingerWu2012,Bandeira2013,Cloninger2024}, which measures
disagreement after translation. We use this
principle to develop a gauge-covariant theory of collective comparison,
relating the geometry of translations to the features a group can share
and recover.

What the move exposes appears in a round trip. Individual $i$ expresses
$j$'s state in its own terms through a map we call a \emph{translation}, or
\emph{dictionary}, $U_{ij}\colon V_j\to V_i$. Suppose every pair is exactly
reciprocal, so that translating a state to a neighbor and back returns it
unchanged (Fig.~\ref{fig:loop}\textbf{B(i)--(iii)}). Compose the dictionaries
around a closed chain and the result need not be the identity
(Fig.~\ref{fig:loop}\textbf{C}). Every pair returns a state
intact, yet the group can change it: pairwise reversibility does not
guarantee collective compatibility. The accumulated map is the holonomy of
the loop. Private relabelings only conjugate it, so it belongs to the
relations rather than to anyone's naming convention \cite{Gao2021}, and
dictionaries that are all conversions from one shared representation leave
no mismatch at all. The defect has no address in any pair. Its smallest
witness is a closed loop.

Two lines of work bring this question to the surface. One has made internal
representation physically consequential: in collective behaviour
\cite{Sayin2025,SalahshourCouzin2025,BastienRomanczuk2020} and evolutionary
game theory \cite{SalahshourPerceptual2025,SalahshourAltruistic2025}.
These models, however, do not treat loop consistency as an independent constraint.
Even when private angular coordinates are explicitly allowed
\cite{SalahshourCouzin2025}, they can nevertheless be related by globally
consistent coordinate conversions, silently importing the inherited
shared-frame perspective. The other line directly confronts this gap by making the
relations themselves explicit: angular synchronization and connection
Laplacians \cite{Singer2011,SingerWu2012,Bandeira2013}, the geometry of
synchronization problems \cite{Gao2021,Cloninger2024}, cellular-sheaf models
in which private opinion spaces are linked by communication maps
\cite{HansenGhrist2021}, and gauge-coupled oscillators whose loop mismatches
constrain collective states \cite{TorresHugas2026}. The two lines meet in
this paper: we ask what translations between internal representations imply
for the features a collective actually encodes. What can a collective hold
in common, after all?

We study fixed networks with reciprocal translations that preserve lengths
and inner products. The dictionaries are specified inputs; how a living pair
learns one, and how an experimenter would identify it, remain open. In this
setting the compatible collective states are the feature patterns left
unchanged by every loop, a characterization that follows from an established
kernel relation for the connection Laplacian \cite{SingerWu2012,Gao2021} and
whose consequences we develop across directional, categorical, and
multicomponent representations. Those consequences are selective. One loop
can reverse an arrow and preserve the axis it lies on, so that the group
agrees on the line and not on which way along it; another can erase a single
categorical distinction and leave the rest available. Whether a mismatch
matters therefore depends on what is represented, and we give the exact
condition under which a loop is invisible in every encoded feature.
Shared-frame comparison is contained within the framework as the globally
consistent case: every dictionary can be made the identity by relabeling
private coordinates, and the linear dynamics introduced in Sec.~\ref{sec:spectrum} becomes
ordinary consensus.
The framework opens the surrounding space of relations whose loop
mismatches no relabeling can remove. We quantify how often independently
and uniformly sampled dictionaries from a finite group close consistently,
how long incompatible features persist, and how uncertainty in individual
dictionaries limits what can be inferred about loop mismatch.

Shared content finally acquires an operational meaning. A message reaches
a receiver who knows every dictionary but not the route it took. The
dictionaries constrain what the message could mean, but do not identify
the transformation it underwent. For our specified random-route model,
with an isotropic Gaussian source and isotropic Gaussian readout noise,
we derive the smallest mean-squared reconstruction error achievable by
any decoder. At fixed signal-to-noise ratio, the fraction of feature
dimensions preserved jointly by all loops determines the long-route
limit. Two networks whose loops have identical rotation angles can
preserve different features, depending on whether their rotation axes
coincide. In this reconstruction task, recording the route removes the
resulting difference in error. The relational geometry of a collective
thus sets both the states it can share and the content it can recover.

\section{Private worlds, translations, and relabeling}
\label{sec:gauge}

\subsection{Setup}

Let a finite set of individuals $i=1,\dots,N$ interact along the edges of a
connected, simple undirected graph $\mathcal{G}$: vertices are individuals,
edges are pairwise relations, and every individual is reachable from every
other. There are no self-edges or
parallel edges \cite{Diestel2017}. Individual $i$ carries a state $a_i$
in a private representation space $V_i\cong\mathbb{C}^{d}$ (or
$\mathbb{R}^{d}$), a vector space with $d$ feature coordinates; the dimension $d$ is the same for all individuals---an
idealization discussed under Scope and next questions in
Sec.~\ref{sec:discussion}.

A feature specifies the kind of content encoded, such as a direction toward
food, an axis without a preferred end, or a contrast between food categories.
We call a set of feature coordinates closed under translation a
\emph{feature sector}: translating any vector in that space keeps it in the
same space.

Consider individuals encoding the direction of food as an arrow in a private
angular frame. At angle $\theta$, its unit-length coordinates are
$(\cos\theta,\sin\theta)$. Translating between individuals rotates these
components while preserving the arrow's length; the directional sector is
closed because every such translation produces another directional arrow.
If an individual encodes only the arrow's axis, without distinguishing its
two ends, suitable coordinates are $(\cos 2\theta,\sin 2\theta)$: they assign
the same value to $\theta$ and $\theta+\pi$ \cite{MardiaJupp1999}. A half-turn reverses the arrow
but leaves its axis unchanged. The same translation can therefore act
differently on different encoded features.

Direction and axis are examples of a larger family. On the circle $S^1$,
the angular sector of order $m$ uses the pair
$(\cos m\theta,\sin m\theta)$, or equivalently $e^{im\theta}$ in complex
notation. Direction has order $m=1$ and axis has order $m=2$.
An angular signal decomposes into Fourier components with these rotation
rules: a rotation through $\alpha$ multiplies the order-$m$ component by
$e^{im\alpha}$; composing translations adds their rotation angles
\cite{BrockerDieck1985}. Supplemental Material,
Sec.~\SIref{sec:s_circle_features}, motivates these coordinates and derives their transformation
rule \cite{SupplementalMaterial}.

\begin{figure*}[!t]
\includegraphics[width=\textwidth]{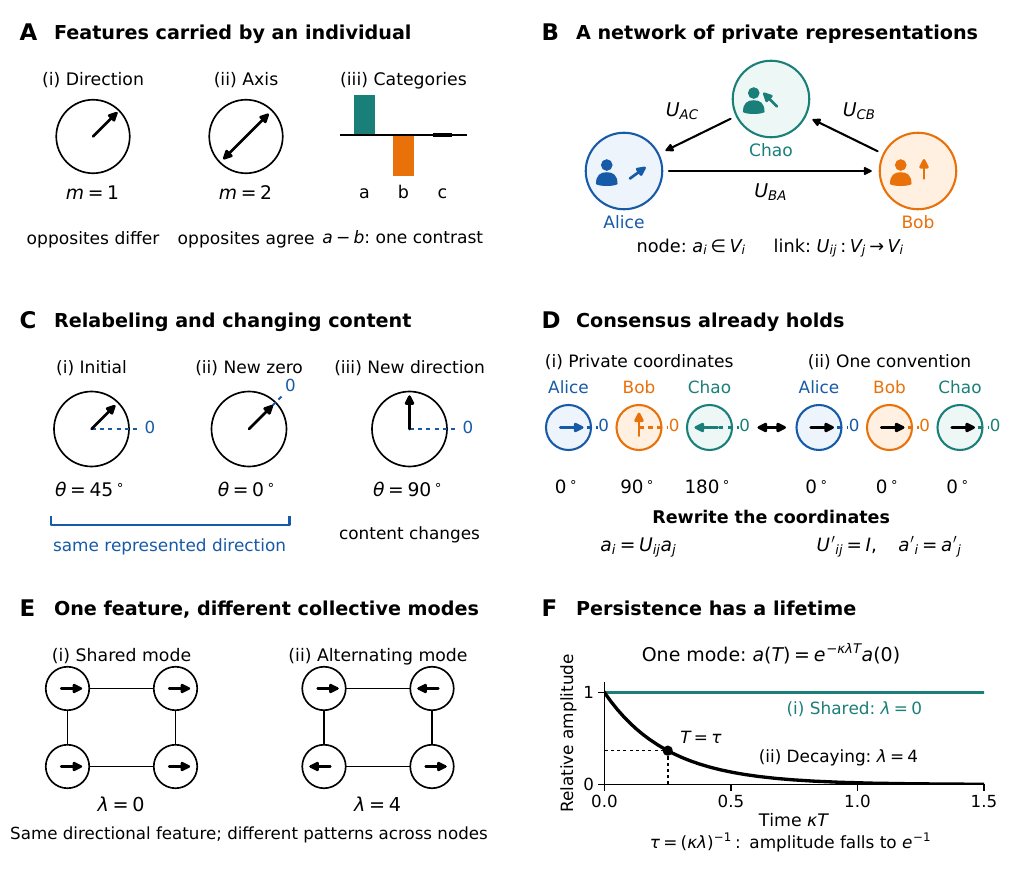}
\caption{\textbf{From private representations to collective patterns.}
\textbf{A(i)--(iii)}~Examples of private representations considered here,
with feature states $a_i\in V_i$: an oriented direction, an unoriented axis, and a
categorical contrast $(1,-1,0)$ on alternatives $a,b,c$.
Direction and axis have angular orders $m=1$ and $m=2$.
Categorical translations permute coordinates within the centered-contrast
space; the displayed contrast is one vector in that space.
\textbf{B}~The graph connects individuals, each carrying a state $a_i$ in
its own space $V_i$. The dictionary $U_{ij}$ translates from $j$ to $i$;
reverse dictionaries are inverses, and only one direction per edge is drawn.
Colored circles enclose private representations;
their arrows use the same individual colors.
\textbf{C(i)--(ii)}~Changing the private angular zero (dashed radius) changes
the \textit{representation} angle from $45^\circ$ to $0^\circ$ while keeping the
represented direction (solid arrow) fixed.
\textbf{C(iii)}~With the original zero retained, turning the arrow changes the
content (the represented direction).
\textbf{D(i)}~Each local chart is drawn with its dashed zero ray pointing
right. Colored arrows show the private coordinate vectors at
$0^\circ,90^\circ,180^\circ$, not physical headings on a common compass.
These distinct representations encode the same eastward food direction:
$a_i=R(\alpha_i)(1,0)^{\mathsf T}$,
$(\alpha_A,\alpha_B,\alpha_C)=(0,\pi/2,\pi)$, and
$U_{ij}=R(\alpha_i-\alpha_j)$.
Their readouts $R(-\alpha_i)a_i$ agree, and already $a_i=U_{ij}a_j$.
\textbf{D(ii)}~Rewriting with $R(-\alpha_i)$ at each node makes every
dictionary the identity and every coordinate angle zero; black arrows show
the states in this common frame. The physical
direction and agreement are unchanged; this is a change of description,
with no time evolution. Both descriptions are flat.
\textbf{E(i)--(ii)}~A four-node cycle with unit edge weights and identity
dictionaries supports a uniform mode with $\lambda=0$ and an alternating
mode with $\lambda=4$. Both use the same directional feature; their patterns
across individuals differ.
\textbf{F}~Under $\dot a=-\kappa L_Ua$, these modes retain relative
amplitudes $1$ and $e^{-4\kappa T}$.
The decaying amplitude reaches $e^{-1}$ at its lifetime
$\tau=1/(4\kappa)$. Arrows represent feature amplitudes;
this linear law does not, and need not, preserve unit headings.}
\label{fig:conceptguide}
\end{figure*}

For categorical content, translations permute labels. On a six-alternative
space we study the \emph{centered contrasts}, whose six coordinates sum to
zero: they encode differences between alternatives after removing their
common baseline. Permuting coordinates preserves this five-dimensional
space \cite{Serre1977}. For directions on the sphere $S^2$, we study angular sectors whose
coordinates mix only within the same sector under a three-dimensional
rotation. These include the ordinary vector sector and higher-order angular
patterns \cite{BrockerDieck1985,Olver2010Handbook}. Supplemental Material, Secs.~\SIref{sec:s_categorical_features} and \SIref{sec:s_spherical_features}, explains these choices
and their transformation rules \cite{SupplementalMaterial}.
Figure~\ref{fig:conceptguide}\textbf{A} illustrates direction, axis, and
categorical content; Fig.~\ref{fig:conceptguide}\textbf{B} shows their
placement on an interaction graph.

More generally, translations belong to a finite or compact matrix group
$G$, a set closed under composition and inversion. A unitary representation
$\rho_r$ assigns a length-preserving matrix to each $g\in G$ on a space
$V^r$ of dimension $d_r$, with
$\rho_r(g_1g_2)=\rho_r(g_1)\rho_r(g_2)$. For compact groups we use
continuous representations \cite{Serre1977,BrockerDieck1985}.
The index $r$ runs over a specified set $\mathcal R$ of encoded feature
sectors. The angular and categorical sectors above are worked examples;
the general results admit other sectors under the same assumptions.
The encoded subset of sectors and the pairwise dictionaries are model inputs,
to be specified or measured in an application. The framework determines
their compatibility; it does not infer which features individuals encode.

A message is a particular value of the encoded content. Thus $(x,y,z)$ and
$(x,0,z)$ can be two messages in the same three-dimensional space; a zero
component does not remove a feature coordinate. A collective pattern assigns
a message to every individual. We use \emph{collective mode} in
Sec.~\ref{sec:spectrum} for a pattern that changes only in amplitude under
the specified linear relaxation. One feature sector can support many such
network patterns and infinitely many message values.

Each edge $\{i,j\}$ carries a group-valued \emph{translation} $U_{ij}\in G$.
Its matrix $\rho_r(U_{ij})\colon V_j^r\to V_i^r$ is the rule by which $i$
interprets $j$'s state in sector $r$. When discussing one sector we suppress
$\rho_r$ and write $U_{ij}$ for this matrix. The distinction matters when
a nonidentity group element acts as the identity on an encoded sector.
We assume the translations are unitary, meaning they preserve lengths and
inner products, and \emph{reciprocal},
\begin{equation}
U_{ji}\;=\;U_{ij}^{-1}\;=\;U_{ij}^{\dagger},
\label{eq:reciprocity}
\end{equation}
where $\dagger$ denotes the conjugate transpose and $I$ below denotes the
identity map. Translating from $j$ to $i$ and back returns every state exactly.
This is internal consistency of the pairwise dictionaries; accuracy against
an external stimulus requires an independent measurement
\cite{Salahshour2019,SalahshourCommunication2019,SalahshourPrivateNoise2026}. Reciprocity of the
\emph{dictionary}, Eq.~\eqref{eq:reciprocity}, is a hypothesis about learned
translations and must not be confused with reciprocity of an interaction
\emph{force law}; we return to this distinction in Sec.~\ref{sec:discussion}.
When Eq.~\eqref{eq:reciprocity} fails, for example through independently
learned maps that are not mutual inverses, the framework retains the general directed difference operator, but
the reciprocal-unitary spectral results require a separate analysis
(Remark~\ref{rem:directed}).

Each individual's states can be expressed in a different coordinate convention: a
\emph{relabeling}, also called a gauge transformation, is a choice of
$h_i\in G$ on each $V_i$, represented by $\rho_r(h_i)$ within sector $r$, acting by
\begin{equation}
a_i\mapsto h_i a_i,
\qquad
U_{ij}\mapsto h_i\,U_{ij}\,h_j^{-1}.
\label{eq:gauge}
\end{equation}
Relabeling changes no measurable prediction when states, dictionaries, and
readout rules are transformed together. This freedom of private convention
requires physical statements to be invariant under
Eq.~\eqref{eq:gauge} \cite{Wilson1974,Gao2021}. The relabelings used for
the dynamical results are fixed in time.

In the food-direction example, suppose one individual chooses a different
zero from which to measure angles. The same encoded direction receives different coordinates, although
the direction itself has not changed (Fig.~\ref{fig:loop}\textbf{A}). This is a relabeling, not a turn of the
arrow. The dictionaries connecting that individual to its neighbors must be
rewritten to accommodate the new convention, and the readout must interpret
the new coordinates accordingly. Equation~\eqref{eq:gauge} expresses this
coordinated change of description. Different individuals may choose different
angular zeros without changing any measurable prediction. Requiring the
relabelings to be fixed in time means that each change of angular zero---the
offset specified by $h_i$---remains constant throughout the time evolution. The
offset is fixed, not the encoded direction: individuals may still change their
estimates of where food lies as the dynamics unfolds
(Fig.~\ref{fig:conceptguide}\textbf{C}). Fig.~\ref{fig:conceptguide}\textbf{D--F} previews the shared-frame comparison of
Sec.~\ref{sec:containment}, and the collective modes and lifetimes of
Sec.~\ref{sec:spectrum}.

A \emph{loop} $C$ is a closed walk $i_0\to i_1\to\dots\to i_0$ along edges of
$\mathcal{G}$. Its \emph{holonomy} is the composed translation experienced by
a state carried once around,
\begin{equation}
H_C \;=\; \overleftarrow{\prod_{e\in C}} U_e
\;=\;U_{i_0 i_{k-1}}\cdots U_{i_2 i_1}U_{i_1 i_0}\colon V_{i_0}\to V_{i_0},
\label{eq:holonomy}
\end{equation}
where the arrow records composition in travel order \cite{Gao2021}.
Figure~\ref{fig:loop}\textbf{C} shows three individuals encoding a planar
direction ($V_i\cong\mathbb{R}^2$), with every dictionary applying a
half-turn $R(\pi)=-I_2$. Two half-turns restore the message, so each pair
is reciprocal, yet $H_C=R(\pi)^3=-I_2$: an arrow
sent around the triangle returns reversed. No nonzero directional pattern
can satisfy every translation, because its value at Alice would have to
obey $a_A=-a_A$.

The loop is a route, and its holonomy is a transformation: the same $H_C$
acts on every message $v$, returning $H_Cv$. The individual dictionaries
determine this transformation through Eq.~\eqref{eq:holonomy}. Different messages may therefore return
unchanged or altered under the very same loop.

\subsection{What relabeling can and cannot remove}

\begin{proposition}[Conjugacy invariance]
\label{prop:conjugacy}
Under the relabeling \eqref{eq:gauge}, the holonomy of a loop based at $i_0$
transforms as $H_C\mapsto h_{i_0}H_C\,h_{i_0}^{-1}$. Consequently the
conjugacy class of $H_C$, the matrices related by such a coordinate
change---in particular its spectrum, the eigenvalues counted with multiplicity, and its characters
$\chi_r(H_C)=\tr\rho_r(H_C)$, the sums of diagonal entries in each sector, and the dimension of its fixed space
$\Fix(H_C)=\{v:H_Cv=v\}$---is invariant under all private relabelings.
\end{proposition}

The proof is a telescoping cancellation of the interior gauges
(Supplemental Material, Sec.~\SIref{sec:s2}); the statement itself is the discrete form of a
standard gauge-theory fact \cite{Wilson1974,Kogut1979}. Its physical content
here: whether meaning returns preserved, rotated, or reversed around a loop is
a property of the loop that survives every change of naming convention. We call the
conjugacy-class content of $H_C$ the \emph{relational flux} of the loop, and a
loop with nontrivial flux a \emph{defect}. The connection with gauge physics
is structural: the matrices depend on local conventions, while the loop's
conjugacy class does not. Observable consequences belong to this invariant
relational content.

\begin{proposition}[Dyads are blind]
\label{prop:dyad}
(i) Under reciprocity \eqref{eq:reciprocity}, a dyad, or pair of individuals,
has round trip $U_{ij}U_{ji}=I$ identically and carries no nontrivial
round-trip holonomy.
(ii) For unitary translations, an open-path composite
$P=U_{i_k i_{k-1}}\cdots U_{i_1 i_0}$ transforms as
$P\mapsto h_{i_k}P\,h_{i_0}^{-1}$ with \emph{independent} endpoint gauges, so
relabeling can bring any open-path composite to the identity when its
endpoints are distinct. A nonreciprocal dyad can have a nonidentity round
trip; the first conclusion then does not apply.
\end{proposition}

\begin{proposition}[Trees flatten; $b_1$ loop generators remain]
\label{prop:tree}
A spanning tree connects all $N$ vertices with $N-1$ edges and has no
simple cycle, a route through at least three distinct vertices returning
to its start without revisiting any other vertex \cite{Diestel2017}. On any spanning tree of
$\mathcal{G}$ there is a relabeling bringing
every tree edge to $U_e=I$ (choose $h_i$ to be the inverse of the tree transport
from a root to $i$). After this choice, exactly
$b_1=E-N+1$
fundamental-loop holonomies remain, one per non-tree edge. The integer $b_1$
is the cycle rank, or first Betti number \cite{Diestel2017,Hatcher2002}.
These holonomies generate
$\Hol_{i_0}(U)$, the group of round trips beginning and ending at the chosen
root $i_0$. A common conjugation at the root is still free
\cite{Hatcher2002,Gao2021}.
\end{proposition}

\begin{proposition}[Loops cannot be flattened]
\label{prop:sharp}
If $H_C\neq I$, no relabeling makes the loop trivial: conjugation preserves
the spectrum, and $I$ is the only unitary whose spectrum is $\{1\}$
\cite{HornJohnson2013}.
\end{proposition}

Propositions~\ref{prop:dyad}--\ref{prop:sharp} identify the smallest
geometric obstruction: \emph{on a simple graph with reciprocal
translations, the smallest possible nontrivial loop contains three
individuals}. On an open
path the two loose gauge ends belong to different individuals and can be
chosen independently; on a loop they are the same individual, and only
conjugation remains. A pair or tree therefore supplies no independent closed-loop constraint.
Supplemental Material, Sec.~\SIref{sec:s2}, proves these three propositions;
Fig.~\SIref{fig:s1} illustrates the constructive relabeling and its surviving loop mismatch.

Return to the food-direction example, with dictionaries that rotate the arrow
by a half-turn, $R(\pi)$, in either direction along every pair.
Each \emph{dyad}---a pair such as Alice and Bob---returns the arrow unchanged
on a round trip, because two half-turns make a full turn.
The triangle does not: its \emph{loop holonomy} is the accumulated
return transformation $H_C=R(180^\circ)=-I_2$, written here on the arrow's two
real components. Changing private angular zeros changes the individual
dictionaries but cannot remove this half-turn. Coordinate-equivalent
descriptions of the return transformation form its \emph{conjugacy class};
in this planar example, they all give the same rotation matrix. The
unavoidable half-turn is therefore the loop's \emph{relational flux}, and
the loop is a \emph{defect} because its return transformation is not the
identity. Its spectrum consists of two eigenvalues $-1$, expressing reversal
of both arrow components; its character is their sum, $-2$. Its
\emph{fixed space} contains only the zero vector: no nonzero arrow survives
the journey unchanged. An axis feature, however, does survive, because
reversing its two ends leaves the represented axis intact.

\begin{remark}[Directed translations]
\label{rem:directed}
If reciprocity or unitarity fails, states of zero translated disagreement are
the kernel of the directed difference operator
$(D_Ua)_e=a_{t(e)}-U_e a_{s(e)}$, as in the difference-operator construction for sheaves \cite{HansenGhrist2021}; the results below are not claimed. Here $s(e)$ and $t(e)$ denote the source and target of an oriented edge.
Applications must measure the accuracy of Eq.~\eqref{eq:reciprocity};
$D_U$ remains the relevant consistency operator when it fails. Nothing in this paper depends on
reciprocity holding in any particular living system.
\end{remark}

\section{Flatness under independent uniform translations}
\label{sec:counting}

A translation system is a \emph{connection} \cite{Gao2021}.
We call it \emph{globally flat}, or briefly flat, when every loop
holonomy is the identity. By Proposition~\ref{prop:tree}, flatness is
equivalent to the existence of a relabeling making every edge the identity,
i.e., to what we call node-factorized translations $U_{ij}=h_ih_j^{-1}$.

\begin{theorem}[Flat fraction]
\label{thm:counting}
Let $\mathcal{G}$ be connected with $E$ edges and $N$ nodes, and let each edge
translation be drawn independently and uniformly from a finite group $G$.
Then
\begin{equation}
\Pr[\text{connection flat}]\;=\;|G|^{-b_1},
\qquad b_1=E-N+1 .
\label{eq:counting}
\end{equation}
\end{theorem}

\begin{figure}[!t]
\includegraphics[width=\columnwidth]{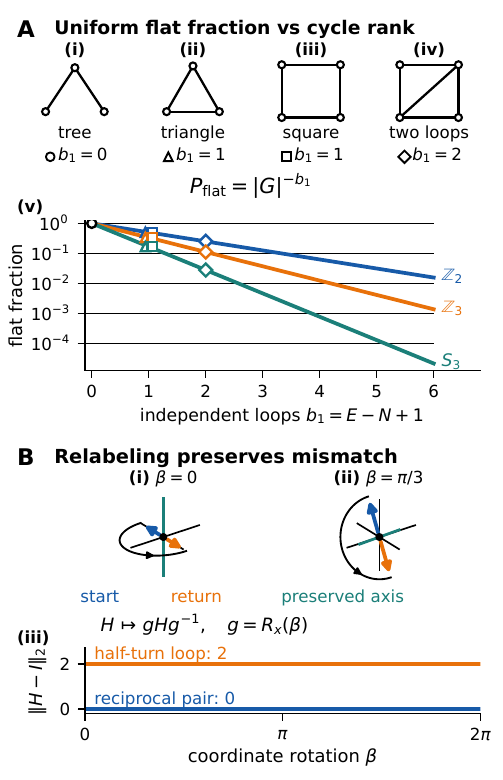}
\caption{\textbf{Independent loops control uniform flatness; private coordinates do not remove loop mismatch.}
\textbf{A(i)--(iv)}~A tree, triangle, square, and four-node two-loop network
have cycle ranks $b_1=0,1,1,2$. Nodes represent individuals, and edges carry
reciprocal dictionaries. Their marker shapes identify the same graphs in
\textbf{A(v)}, where curves give $|G|^{-b_1}$ for independent uniform
dictionaries from $\mathbb Z_2$, $\mathbb Z_3$, and $S_3$, and markers give
exhaustive enumeration. Triangle and square markers are displaced
horizontally by $\mp0.06$ for visibility.
\textbf{B(i)--(ii)}~The same three-dimensional half-turn in two coordinate
conventions related by $g=R_x(\beta)$, with $\beta=0,\pi/3$.
Blue and orange arrows show a test vector and its return.
The curved black arc follows the half-turn between them: it is an
illustrative rotation path.
The teal line is the preserved axis; straight black lines are coordinate
guides. The loop matrix, vector, and axis transform together,
$H\mapsto gHg^{-1}$: no physical intervention occurs.
\textbf{B(iii)}~The operator distance from identity remains $2$ for the
half-turn and $0$ for a reciprocal pair over 201 coordinate rotations
$\beta\in[0,2\pi]$. Supplemental Material, Secs.~\SIref{sec:s2}--\SIref{sec:s3}, gives the calculations.}
\label{fig:counting}
\end{figure}

The proof (Supplemental Material, Sec.~\SIref{sec:s3}) gauges a spanning tree to the identity;
the $b_1$ surviving generators are then independent and uniform, and each must
separately equal the identity. Exhaustive enumeration gives the same exact
fractions over twelve
graph--group cases ($\mathbb{Z}_2$, $\mathbb{Z}_3$, $S_3$ on the triangle,
square, theta graph, and a tree; Fig.~\ref{fig:counting}\textbf{A(i)--(v)});
for trees the
fraction is $1$, as Eq.~\eqref{eq:counting} requires.

\begin{corollary}
\label{cor:720}
For a single triangle of individuals holding six alternatives with learned
permutation dictionaries ($G=S_6$, $b_1=1$), the flat fraction under the
uniform measure is $1/720$.
\end{corollary}

To see the counting directly, let Alice, Bob, and Chao each distinguish six
food categories. Once the Alice--Bob and Bob--Chao dictionaries are fixed,
exactly one of the $6!=720$ possible Chao--Alice dictionaries closes the
triangle consistently for every category. Uniform sampling gives that choice
probability $1/720$. Each additional independent loop imposes another closure
condition, contributing a factor $1/|G|$ in Eq.~\eqref{eq:counting}. The law
counts assignments of dictionaries, not messages: an identity return map
preserves every message at once, however many values it can take. Failure
of this condition need not destroy every shared feature. The next section
asks which features survive when full closure fails.

The uniform measure is a modeling choice. Independent finite-group
translations concentrated near the identity can make flatness arbitrarily
likely. Correlations introduced by learning can also enforce it exactly.
Equation~\eqref{eq:counting} is therefore a uniform reference probability;
it does not predict the outcome of an unmeasured learning process (Supplemental Material,
Sec.~\SIref{sec:s3} and Fig.~\SIref{fig:s2}). The two coordinate views in
Fig.~\ref{fig:counting}\textbf{B(i)--(ii)} depict the same loop mismatch.
Changing coordinates by $R_x(\beta)$ leaves a half-turn loop at distance
two from the identity for every $\beta$, while a reciprocal pair remains
at zero distance (Fig.~\ref{fig:counting}\textbf{B(iii)}).
Unitary conjugation preserves this operator distance
(Supplemental Material, Eq.~(\SIref{eq:s_conjugation_distance})).

\section{The existence theorem}
\label{sec:existence}

Which represented features remain compatible in the presence of loop mismatch? Fix a
feature sector $r$ and give each individual a state $a_i\in V^r$. Let
$w_{ij}=w_{ji}>0$ be interaction weights on the edges. Translated
disagreement across edge $\{i,j\}$ is $a_i-\rho_r(U_{ij})a_j$; summing over
neighbors defines the \emph{connection Laplacian}
\cite{SingerWu2012,Bandeira2013}
\begin{equation}
(L^r_U a)_i \;=\; \sum_{j}w_{ij}\bigl(a_i-\rho_r(U_{ij})\,a_j\bigr).
\label{eq:laplacian}
\end{equation}
Under reciprocity, $L^r_U$ is Hermitian (equal to its conjugate transpose)
and positive semidefinite: its quadratic form is
$a^\dagger L^r_Ua=\tfrac12\sum_{ij}w_{ij}\lVert a_i-\rho_r(U_{ij})a_j\rVert^2\geq0$.
Its kernel, the set of vectors it sends to zero, consists of
\emph{patterns with no translated disagreement anywhere}. We use
\emph{common ground} for this space of compatible patterns. The zero pattern
always belongs to it; a positive kernel dimension counts independently
shareable feature amplitudes. Even a one-dimensional shared space contains
infinitely many values of its surviving amplitude. These are linear features, not necessarily
unit headings or complete probability distributions.

\begin{theorem}[What a collective can hold in common]
\label{thm:kernel}
Let $\mathcal{G}$ be connected with symmetric positive weights, and let the
sector translations be unitary and reciprocal. Then
\begin{equation}
\ker L^r_U \;\cong\; \bigcap_{C\ {\rm based\ at}\ i_0}
\Fix\bigl(\rho_r(H_C)\bigr),
\label{eq:kernel}
\end{equation}
The symbol $\cong$ in Eq.~\eqref{eq:kernel} denotes a one-to-one linear
correspondence between these two spaces. A vector $v$ in the joint fixed space
at the reference individual $i_0$ determines a compatible collective pattern
through $a_i=P_i v$. Here $P_i$ is the product of sector translations along
the unique path from $i_0$ to $i$ in a chosen spanning tree. Conversely, every
compatible pattern determines such a vector through $v=a_{i_0}$. To find the
joint fixed space, it suffices to intersect the fixed spaces of the $b_1$
fundamental-loop generators from Proposition~\ref{prop:tree}: a vector
unchanged by these generators is unchanged by every loop.
\end{theorem}

\emph{A collective can hold in common exactly what all of its loops leave
unchanged.} The proof (Supplemental Material, Sec.~\SIref{sec:s4}) uses positivity to
force agreement across each edge and tree paths to carry a reference value
through the network. This is an established connection between compatible
patterns and holonomy \cite{Gao2021,SingerWu2012,HansenGhrist2021}; its
physical interpretation here is a count of shared represented features.
Let $\mathcal{H}_r=\overline{\langle\rho_r(H_C)\rangle}$ contain all products
of based loop matrices, their inverses, and their limits. It is a compact
matrix group. Averaging its matrices with the normalized invariant
probability measure $\mu$ gives the projector
$P_{\mathrm{share}}=\int_{\mathcal{H}_r}h\,d\mu(h)$ and dimension
$d^{\,r}_{\mathrm{share}}=\int_{\mathcal{H}_r}\tr h\,d\mu(h)$
\cite{BrockerDieck1985}.
Supplemental Material, Sec.~\SIref{sec:s4}, derives this projector.

Two examples show why the intersection of fixed spaces matters.

\begin{proposition}[Non-flat but shareable]
\label{prop:guard1}
On a triangle of three-dimensional rotational representations with loop
holonomy $R_z(\pi)$ (a half-turn about the $z$ axis), the connection is not
flat, yet $\ker L_U$ is one-dimensional: the $z$ component is held in common
while no nonzero transverse pattern is exactly compatible. Thus
$\lambda_{\min}=0$ does not imply flatness.
\end{proposition}

Here the half-turn is only illustrative: any nonidentity rotation about the
$z$ axis has the same one-dimensional fixed space in
the three-dimensional vector sector.

\begin{proposition}[Two-loop erasure]
\label{prop:guard2}
Two based loops---realized on a four-node theta
graph---with holonomies $R_z(\pi/2)$ and $R_x(\pi/2)$ each fix an
axis, but their joint fixed space is $\{0\}$: noncommuting defects can erase
every shareable vector in a sector even though each loop, alone, would spare
one.
\end{proposition}

Figure~\SIref{fig:s3} shows the fixed spaces and their intersection; Supplemental
Material, Sec.~\SIref{sec:s4}, proves both propositions by finding these spaces.
The mechanism is the absence of a
direction preserved by both loops. Noncommutativity is not necessary:
half-turns about the $x$ and $z$ axes commute and also have joint fixed
space $\{0\}$. Separate loop spectra therefore need not determine common
ground on a multi-loop graph.

Finally, one exactly solvable sector ties this framework to signed-network
theory and will anchor the containment argument:

\begin{proposition}[Even and odd features under sign reversal]
\label{prop:z2}
On a connected graph with symmetric positive edge weights, let each
reciprocal dictionary be a sign $s_{ij}\in\{+1,-1\}$, acting on feature
sector $m$ as $s_{ij}^m$. For even $m$, every dictionary acts as the
identity because $s_{ij}^m=1$; the connection Laplacian therefore reduces
to the ordinary graph Laplacian. For odd $m$, $s_{ij}^m=s_{ij}$, so
reversals remain visible and the operator is the signed Laplacian.

A \emph{negative cycle} is a closed cycle containing an odd number of
reversing edges, giving a return transformation of $-1$ in every odd
sector. If such a cycle exists, no nonzero compatible pattern survives
in any odd sector, and its smallest Laplacian eigenvalue is strictly
positive. Changing private sign conventions, while rewriting the
dictionaries consistently, leaves all eigenvalues unchanged: the
relabeling is \emph{isospectral} \cite{Harary1953,Altafini2013}.
\end{proposition}

For the directional example, this means that a loop can obstruct sharing
an arrow ($m=1$) while leaving its unoriented axis ($m=2$) fully shareable.
The sign parity and the fixed-space criterion in Eq.~\eqref{eq:kernel}
give the result; Supplemental Material, Sec.~\SIref{sec:s6}, supplies the proof.

\section{Functional flatness: when a defect is invisible}
\label{sec:functional}

A loop mismatch can be invisible to the features a collective represents.
A half-turn reverses the food-direction arrow while preserving its axis.
If individuals encode only the axis, this mismatch changes none of their
represented content. We call full sharing of all encoded features
\emph{functional flatness}.

Let $\mathcal{R}$ denote the feature sectors the individuals encode, with
representations $\rho_r$. We call the transformations that leave every
value in every encoded sector unchanged the \emph{blind subgroup},
\begin{equation}
K_{\mathcal{R}}\;=\;\bigcap_{r\in\mathcal{R}}\ker\rho_r .
\label{eq:blind}
\end{equation}
Here $\ker\rho_r=\{g\in G:\rho_r(g)=I\}$: it contains transformations
whose action on sector $r$ is the identity. Taking the intersection selects
those invisible to all encoded sectors.

\begin{theorem}[Functional flatness]
\label{thm:functional}
Under the hypotheses of Theorem~\ref{thm:kernel} in each sector, the
collective has full common ground in every represented sector,
$\dim\ker L_U^r=d_r$, if and only if
\begin{equation}
\Hol_{i_0}(U)\;\subseteq\;K_{\mathcal{R}} .
\label{eq:functional}
\end{equation}
Thus, every return transformation generated by the network's loops must
leave every encoded feature unchanged.
\end{theorem}

The proof applies Eq.~\eqref{eq:kernel} in each encoded sector
(Supplemental Material, Sec.~\SIref{sec:s5}).

Full common ground means that any feature value at the reference individual
can be extended into a compatible pattern across the network. Geometric
flatness guarantees this because every loop returns the identity.
Functional flatness also allows nonidentity returns, provided their action
is invisible in all encoded sectors. The two conditions coincide when the
sectors are \emph{jointly faithful}: together, they distinguish every
nonidentity transformation from the identity \cite{Serre1977}.

Transformations with identical effects on all encoded features are
identified in the \emph{quotient group} $G/K_{\mathcal{R}}$
\cite{DummitFoote2004}. For an axis-only representation, rotations differing
by a half-turn have the same effect and belong to the same equivalence
class. This quotient describes the distinctions between translations that
remain visible through the encoded features.

\begin{proposition}[Angular features: the greatest-common-divisor rule]
\label{prop:gcd}
Let $M$ be a nonempty finite set of encoded, nonzero angular
orders. A loop rotation through $\Omega$ acts on sector $m$ by
$e^{im\Omega}$, so it is invisible to that sector when $e^{im\Omega}=1$.
Writing $q=\gcd\{|m|:m\in M\}$ for the greatest common divisor of the
encoded orders, their blind subgroup is $K_M=\mathbb{Z}_q\subset U(1)$:
rotations by integer multiples of $2\pi/q$.
\end{proposition}

For axes alone, $M=\{2\}$, a half-turn is invisible. Adding direction gives
$M=\{1,2\}$, whose greatest common divisor is one. Only the identity
rotation is then invisible to both features, and a half-turn obstructs full
sharing.

\begin{proposition}[Categorical features: cycles and orbits]
\label{prop:perm}
Consider $K$ categories with permutation dictionaries. Their centered
contrast space, $\mathbf{1}^\perp\subset\mathbb{R}^K$, consists of vectors
whose components sum to zero. A loop permutation $\pi$ preserves
$c(\pi)-1$ independent contrasts, where $c(\pi)$ counts its disjoint
permutation cycles, including categories left fixed. These cycles describe
exchanges of category labels.

For several loops, categories form \emph{orbits}: groups of labels that
repeated loop translations can exchange with one another. A shared
contrast must have equal values within each orbit. Its dimension is
therefore the number of orbits minus one, with the subtraction accounting
for the zero-sum constraint.
\end{proposition}

For six food categories, a loop that swaps $A$ and $B$ reverses the contrast
$q_{AB}=(e_A-e_B)/\sqrt2$, where $e_A$ denotes unit weight on category $A$.
Four independent contrasts remain unchanged. Compatibility requires equal
values for $A$ and $B$, while allowing distinctions between their combined
category and the other alternatives.

\begin{figure*}
\includegraphics[width=\textwidth]{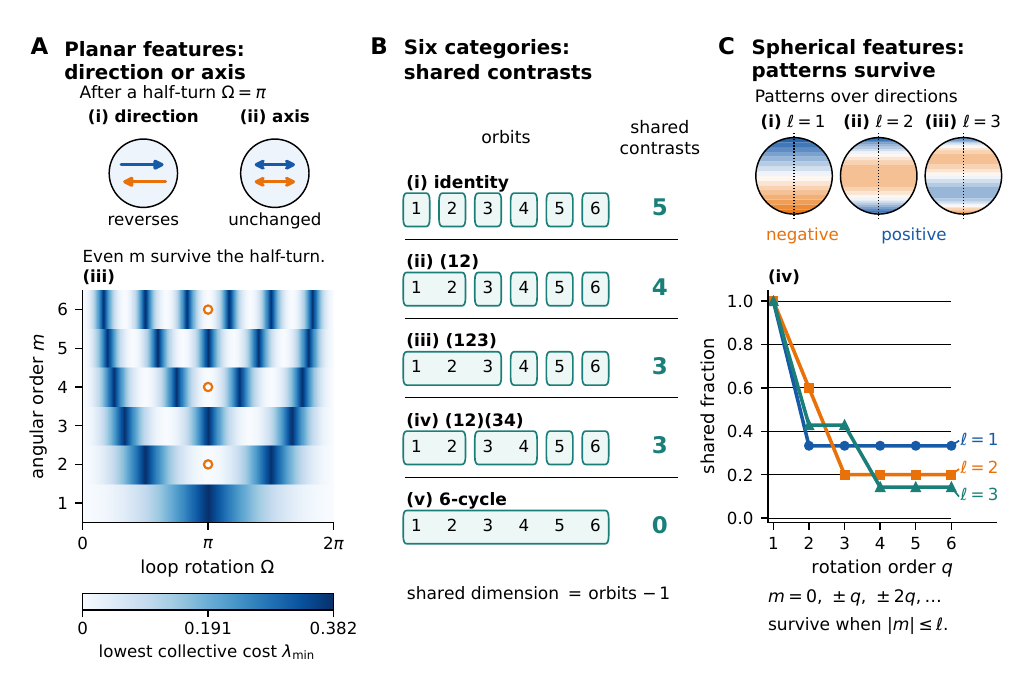}
\caption{\textbf{A loop preserves different content in different representations.}
\textbf{A(i)--(ii)}~A half-turn reverses an oriented direction ($m=1$)
but leaves an unoriented axis ($m=2$) unchanged. Blue and orange denote
initial and returned features; their separation is for visibility.
\textbf{A(iii)}~For a five-node cycle with unit weights and loop rotation
$\Omega$, color gives the lowest connection-Laplacian eigenvalue for each
angular sector of order $m$. Zero cost means an exactly compatible collective
mode; orange-outlined markers identify the even orders surviving at
$\Omega=\pi$. The scale spans the analytic range
$0\leq\lambda_{\min}\leq2[1-\cos(\pi/5)]$.
\textbf{B(i)--(v)}~Six-category permutations: identity, $(12)$, $(123)$,
$(12)(34)$, and a six-cycle. Each teal capsule groups one permutation
orbit: a preserved contrast assigns equal amplitudes within it.
Removing the all-equal component leaves $5,4,3,3,0$ independent centered
contrasts, respectively.
\textbf{C(i)--(iii)}~Example features for three-dimensional directions:
each sphere point specifies a direction, and color gives its signed
feature value. The three patterns are
$z$, $(3z^2-1)/2$, and $(5z^3-3z)/2$, where $z=\cos\theta$ and $\theta$
is the angle from the dotted vertical axis. These are the Legendre
polynomials $P_\ell(z)$ for $\ell=1,2,3$, proportional to the standard
spherical functions $Y_{\ell0}$ \cite{Olver2010Handbook}.
Each is one basis pattern in a $(2\ell+1)$-dimensional feature sector
closed under rotations (SI \SIref{sec:s_spherical_features}). Blue and orange denote positive and
negative values on a common $[-1,1]$ scale. These pictured patterns survive
every rotation about the vertical axis.
\textbf{C(iv)}~A loop rotation $2\pi/q$ preserves spherical orders $m$
divisible by $q$, giving shared fraction
$[2\lfloor\ell/q\rfloor+1]/(2\ell+1)$. Nonzero surviving orders add
features not pictured in \textbf{C(i)--(iii)}. Curves connect exact
integer-$q$ values for visual guidance.
Supplemental Material, Secs.~\SIref{sec:s5} and \SIref{sec:s7}, gives the calculations.}
\label{fig:worlds}
\end{figure*}

Fig.~\ref{fig:worlds}\textbf{A(i)--(iii)} compares directional and
axis features, Fig.~\ref{fig:worlds}\textbf{B(i)--(v)} shows the categorical distinctions
preserved by different permutations, and Fig.~\ref{fig:worlds}\textbf{C(i)--(iv)} extends the
same criterion to spherical features. Supplemental Material, Sec.~\SIref{sec:s5},
proves Propositions~\ref{prop:gcd} and \ref{prop:perm};
Sec.~\SIref{sec:s_spherical_features} derives the spherical fixed dimensions.

Representing fewer features can make additional transformations invisible.
Let an onto linear map $Q\colon V_F\to V_Q$ discard some already encoded feature coordinates
while respecting translations, $Q\rho_F(g)=\rho_Q(g)Q$. This condition
is the standard intertwining relation \cite{BrockerDieck1985}: it means
that translating and discarding features give the same final coarse
representation in either order. Then $K_F\subseteq K_Q$: any
transformation acting as the identity on the detailed representation also
acts as the identity on every coarse value, because $Q$ is onto. The
coarser representation can therefore hide a loop mismatch without changing
the underlying dictionaries or introducing additional coordinate
relabelings. This linear coarsening does not turn a direction into an axis;
that construction is nonlinear. It can, for example, discard the direction
sector from a representation that already encodes both direction and axis.

\section{Containment: the flat sector of collective models}
\label{sec:containment}

\noindent\begin{minipage}{\columnwidth}
\begin{proposition}[Common-frame comparisons sit at $U=I$]
\label{prop:containment}
Direct comparison of states already expressed in one shared
frame corresponds to $U_{ij}=I$. Its gauge-equivalent descriptions have
$U_{ij}=h_ih_j^{-1}$ and form exactly the globally flat sector of the
underlying group-valued dictionaries. The
linear dynamics of Sec.~\ref{sec:spectrum} then reduces to ordinary
continuous-time consensus.
Signed-network consensus \cite{Altafini2013} is the $\mathbb{Z}_2$ case:
Harary's structural balance \cite{Harary1953}, in which every cycle has
positive sign product, is flatness.
\end{proposition}
\end{minipage}

Substituting the factorized dictionaries into Eq.~\eqref{eq:laplacian}
and using Proposition~\ref{prop:tree} gives the containment; Supplemental
Material, Sec.~\SIref{sec:s6}, gives the proof, including the signed case.
If the encoded sectors are not jointly faithful, functional flatness is
already sufficient for ordinary comparison within those sectors: the
underlying dictionaries can still have loops in the blind subgroup.

Figure~\ref{fig:conceptguide}\textbf{D(i)--(ii)} makes this containment
explicit: different private vectors can describe the same compatible state,
and become equal when every dictionary is made the identity.
Common-frame comparison also appears in DeGroot-type and bounded-confidence
models \cite{DeGroot1974,HegselmannKrause2002}, where opinions are compared
on a shared scale. This containment concerns their comparison structure.
The uniform flat fraction in Eq.~\eqref{eq:counting} provides a reference
for independently sampled dictionaries; it does not measure the validity
of those models. Allowing nontrivial loop transformations exposes
feature-dependent compatibility: Eq.~\eqref{eq:kernel} determines which
features can be shared, as illustrated in Fig.~\ref{fig:worlds}. Internal
representations already appear in collective sensing and language evolution
\cite{Salahshour2019,SalahshourCommunication2019,SalahshourLanguage2020,NowakPlotkinKrakauer1999,TrapaNowak2000}.
Their induced comparison maps lie in the flat sector when each private
code is related to one common content space by mutually inverse, linear isometric
encoding and decoding maps. For collective motion, different private angular
zeros likewise remain compatible with shared-frame comparison when their
coordinate conversions cancel around loops
\cite{Vicsek1995,SalahshourCouzin2025}. These correspondences identify a
common comparison structure without establishing equivalence between the
full models.

\section{The spectrum of common ground}
\label{sec:spectrum}

So far we have considered a static problem: which collective patterns
have no translated disagreement anywhere? To study how a collective
approaches such patterns, we introduce a simple local relaxation rule:
each individual adjusts its feature state toward its neighbors' states,
translated into its own coordinates. This is a simple, linear consensus
dynamics with translated comparisons. Summing these weighted adjustments
gives the negative of the connection Laplacian in Eq.~\eqref{eq:laplacian},
and hence the linear dynamics $\dot a=-\kappa L_U^r a$, with $\kappa>0$
setting the time scale \cite{OlfatiSaberMurray2004,HansenGhrist2021}.
This rule decreases the quadratic disagreement and leaves the common-ground
patterns in Eq.~\eqref{eq:kernel} unchanged. Under this dynamics, exact
sharing is the $T\to\infty$ limit of a finite-time question: which
features fade slowly enough to decide with? The dynamics acts on feature
amplitudes; it need not preserve unit headings or the constraints of a
probability density. The moving-particle Vicsek model and Toner--Tu
continuum theory \cite{Vicsek1995,TonerTu1995}, as well as visual-field
and sensing-based models
\cite{BastienRomanczuk2020,Berdahl2013,Salahshour2019}, remain outside
this fixed-graph linear dynamics. Persistence under constrained or richer
intrinsic dynamics requires additional analysis.

\begin{figure*}[t!]
\includegraphics[width=\textwidth]{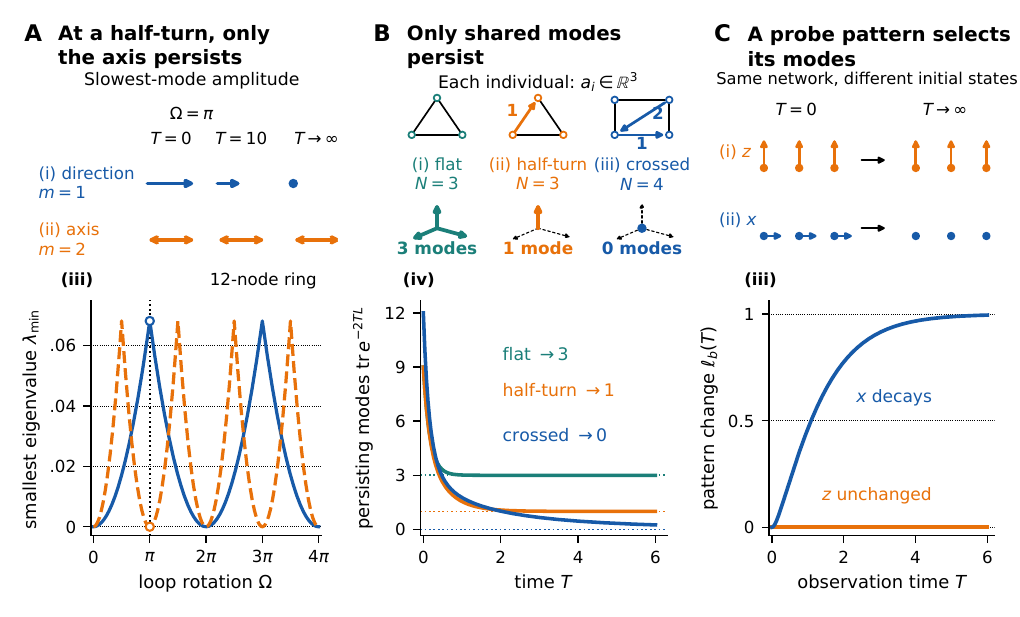}
\caption{\textbf{Represented features determine persistence.}
\textbf{A(i)}~On an $n=12$ ring at loop rotation $\Omega=\pi$, a direction
initialized in its slowest collective mode has normalized amplitude
$e^{-2[1-\cos(\pi/12)]T}$. \textbf{A(ii)}~The corresponding axis amplitude
stays one. Both rows show $T=0$, $10$, and $T\to\infty$; arrow lengths
encode feature amplitudes. Blue denotes direction
($m=1$); orange denotes axis ($m=2$).
\textbf{A(iii)}~The smallest Laplacian eigenvalue against loop rotation.
Direction recurs with period $2\pi$; axis recurs with period $\pi$ and
retains a zero mode at the half-turn.
\textbf{B(i)--(iii)}~Each individual carries a three-component directional
state $a_i=(a_{ix},a_{iy},a_{iz})\in\mathbb R^3$ of variable length.
Black edges carry $I_3$; colored, numbered edges carry the stated rotations
along their arrows and inverse rotations in reverse.
\textbf{B(i)}~Identity translations let the three individuals share any vector.
\textbf{B(ii)}~Edge 1 carries
$R_z(\pi)$, so a trip around the triangle makes a half-turn about $z$;
only the $z$ component can be shared. \textbf{B(iii)}~Edges 1 and 2 carry
$R_z(\pi/2)$ and $R_x(\pi/2)$. Their loops preserve different axes;
only zero is common to both. Below, colored axes show shared components,
dashed black axes guide coordinates, and the dot denotes zero.
\textbf{B(iv)}~The heat trace $\operatorname{tr}e^{-2\kappa T L_U}$ sums
the squared amplitudes retained by all collective modes. At $T=0$,
each mode contributes one: $3N=9$ for either triangle and $3N=12$ for
the four-node graph. Decaying modes contribute less, leaving $3$, $1$,
or $0$ shared modes at long times.
\textbf{C(i)--(ii)}~Initial patterns on the triangle in \textbf{B(ii)}.
Each tuple entry is one individual's full vector:
\textbf{C(i)}~$b_z=(e_z,e_z,e_z)$ gives everyone $e_z=(0,0,1)$ and remains
unchanged; \textbf{C(ii)}~$b_x=(e_x,e_x,e_x)$ instead gives everyone
$e_x=(1,0,0)$ and decays to zero. Dots mark individuals; arrows show
states before and after relaxation.
\textbf{C(iii)}~The normalized squared pattern change $\ell_b(T)$ stays
zero for $b_z$ and tends to one for $b_x$.
All edge weights and $\kappa$ equal one. Supplemental Material,
Secs.~\SIref{sec:s7}--\SIref{sec:s8}, defines the probes and calculations.}
\label{fig:spectrum}
\end{figure*}

\subsection{A worked example: a cycle graph}

To calculate how loop mismatch changes collective relaxation, consider
an equal-weight ring of $n$ individuals, each linked to its two neighbors.
This graph admits an exact spectrum for any feature sector with reciprocal
unitary translations, including directional, categorical, and spherical
representations. The ring specifies the interaction network; it does not
restrict the kind of content represented. Supplemental Material,
Sec.~\SIref{sec:s7}, gives the derivation.

\begin{theorem}[Ring spectrum]
\label{thm:ring}
On an $n$-cycle with equal weights $w$ and reciprocal unitary translations
whose loop holonomy acts in sector $r$ with eigenphases
$\{\phi_{r,a}\}_{a=1}^{d_r}$, the spectrum of $L^r_U$ is exactly
\begin{equation}
\lambda_{r,a,k}\;=\;2w\Bigl[1-\cos\Bigl(\frac{2\pi k+\phi_{r,a}}{n}\Bigr)\Bigr],
\qquad k=0,\dots,n-1 .
\label{eq:ring}
\end{equation}
\end{theorem}

The index $k$ labels a collective mode---a pattern of feature amplitudes
across the $n$ individuals. The index $r$ specifies the encoded sector,
and $a$ labels an eigenvalue $e^{i\phi_{r,a}}$ of its loop-return
transformation. The eigenphase need not describe a physical rotation:
a categorical dictionary that swaps two labels reverses their contrast,
giving eigenvalue $-1$ and phase $\pi$, while unchanged contrasts have
phase zero. Equation~\eqref{eq:ring} applies whether or not the sector
contains nonzero shared features; its values depend on how the loop acts
on that sector. Equation~\eqref{eq:laplacian} takes the encoded features
as given, and the ring theorem determines their collective modes and
relaxation rates $\kappa\lambda_{r,a,k}$.
Figure~\ref{fig:conceptguide}\textbf{E(i)--(ii)} shows two such patterns
using the same directional feature.

For a nonzero angular order $m$, a loop rotation through
$\Omega$ gives $\phi=m\Omega$ modulo $2\pi$. Here $m$ labels content
within each individual's representation, while $k$ labels its pattern
across the group. Substitution into Eq.~\eqref{eq:ring} gives two sharp
signatures (Fig.~\ref{fig:spectrum}\textbf{A(i)--(iii)};
Supplemental Material, Sec.~\SIref{sec:s7}):

\begin{proposition}[Recurrence]
\label{prop:recurrence}
The full spectral set of angular sector $m$ is invariant under
$\Omega\to\Omega+2\pi/|m|$; a shift by half that period is generically not a
recurrence. The axis sector repeats twice as often as the direction sector.
\end{proposition}

\noindent\begin{minipage}{\columnwidth}
\begin{proposition}[Selectivity at half flux]
\label{prop:selectivity}
At $\Omega=\pi$ the direction sector has the smallest eigenvalue
$\lambda_{\min}=2w[1-\cos(\pi/n)]>0$ while the axis sector retains an exact
zero mode: the ring loses ``north rather than south'' and keeps ``the
north--south axis.''
\end{proposition}
\end{minipage}

The loop mismatch shifts the spectrum although it is removable on every open
path (Proposition~\ref{prop:dyad}); at the level of
Eqs.~\eqref{eq:holonomy} and \eqref{eq:ring} this is the same algebra as the
Aharonov--Bohm effect \cite{AharonovBohm1959}, with $e^{im\alpha_{ij}}$
playing the role of a Peierls factor \cite{Hofstadter1976} and the fingerprint of
Sec.~\ref{sec:measure} that of a Wilson loop \cite{Wilson1974,Kogut1979}.
These are exact correspondences of formulas, with the stated scope, and
nothing more: $\kappa\lambda$ is a relaxation rate, and no quantum
mechanics is implied.

\subsection{Finite-time persistence of collective patterns}

Exact compatibility determines what survives indefinitely. At a finite
observation time, incompatible patterns may also remain appreciable.
We therefore ask two questions: how much of the collective's space of
possible patterns persists, and how much does a particular pattern change?
These questions apply to any fixed network under the linear relaxation
introduced above, not only to the ring.

The solution is $a(T)=e^{-\kappa T L_U}a(0)$ \cite{Higham2008}. The matrix exponential is
the time-evolution operator, conventionally called the \emph{heat kernel}
\cite{Tsitsulin2018}: it carries the initial pattern forward by time $T$.
A collective mode with Laplacian eigenvalue $\lambda_\nu$ retains the
fraction $e^{-\kappa T\lambda_\nu}$ of its initial amplitude. Thus
zero-eigenvalue modes remain unchanged, whereas positive-eigenvalue modes
decay. For $\lambda_\nu>0$, the amplitude lifetime
$\tau_\nu=1/(\kappa\lambda_\nu)$ is the time to retain $e^{-1}$ of the
initial amplitude (Fig.~\ref{fig:conceptguide}\textbf{F}).

To examine particular content, we specify a nonzero initial pattern $b$:
the feature amplitudes assigned to each individual. We call this a
\emph{probe pattern}. It can combine several collective modes rather than
coincide with a single one. Its evolved value is
$b(T)=e^{-\kappa T L_U}b$.

\begin{proposition}[Finite-time persistence]
\label{prop:finitetime}
Under $\dot a=-\kappa L_Ua$, with $\kappa>0$:

(i) Overall persistence is measured by the heat trace
\[
\mathcal D(T)=\tr e^{-2\kappa T L_U}
=\sum_\nu e^{-2\kappa T\lambda_\nu},
\]
where the eigenvalues are counted with multiplicity. Each term is the
remaining squared amplitude of a mode initialized with unit amplitude;
this explains the factor $2$. The sum decreases from the total number
of modes to $d_{\mathrm{share}}$, the dimension of the exactly shared
space. At finite time it is generally noninteger: it is a weighted
count of persisting modes, including transient disagreement.

(ii) The normalized squared change of a probe pattern is
\[
\ell_b(T)=\frac{\lVert b(T)-b\rVert^2}{\lVert b\rVert^2}.
\]
It starts at zero and increases towards
\[
\ell_b(\infty)
=1-\frac{\lVert P_{\ker L_U}b\rVert^2}{\lVert b\rVert^2}.
\]
Here $P_{\ker L_U}b$ is the component of the initial pattern belonging
to the compatible space in Eq.~\eqref{eq:kernel}. A fully compatible
pattern remains unchanged; a pattern with no compatible component
decays completely, giving $\ell_b(T)\to1$.
\end{proposition}

Expanding the initial pattern in Laplacian eigenvectors gives both
statements (Supplemental Material, Sec.~\SIref{sec:s8}).

The networks in Fig.~\ref{fig:spectrum}\textbf{B(i)--(iii)} preserve
three, one, and zero components; their heat traces approach those dimensions
in Fig.~\ref{fig:spectrum}\textbf{B(iv)}. Within the same half-turn network, a probe aligned with
the preserved axis remains unchanged (Fig.~\ref{fig:spectrum}\textbf{C(i)}), while a transverse
probe decays (Fig.~\ref{fig:spectrum}\textbf{C(ii)}), giving different pattern changes
(Fig.~\ref{fig:spectrum}\textbf{C(iii)}).
Persistence therefore depends both on the network's translations and on
the content being tested.

Decomposing $b$ into collective modes identifies which decay rates
contribute to its change. The same modal weights determine its overlap
with its initial pattern and, under the driven dynamics specified in
Supplemental Material, Sec.~\SIref{sec:s8}, its response to oscillatory forcing
(Fig.~\SIref{fig:s7}). Unequal feature strengths also matter: equal shared dimensions
can preserve different fractions of source variation (Fig.~\SIref{fig:source_weighting}).

These quantities describe linear feature persistence, not behavioral
accuracy, which requires an observation and task model. Moreover, the
time-evolution operator is invertible at every finite time: attenuation alone
is not irreversible information loss \cite{Higham2008}.

\section{Reconstructing an original message after unrecorded relaying}
\label{sec:reconstruction}

The structure of common ground determines which features a collective can
share, and the relaxation dynamics describes their persistence. That same
geometry also has operational consequences. Here we derive one: a limit on
recovering a message after it has been relayed through private representations
without a record of its route.

An individual receives a relayed report and knows where it originated,
but not the sequence of translations it underwent. Each intermediary may
have used a reversible dictionary, yet the receiver must account for the
accumulated transformation to recover the original report. Which features
remain recoverable when this translation history is unavailable?

Potential applications include directional reports or categorical feature
amplitudes relayed through learned private codes, where the translations
and source statistics satisfy the model's assumptions. The receiver may
need the original direction, a particular contrast, or only a feature
preserved across different interpretations. This task quantifies both
feature protection and the benefit of retaining translation records.
It assumes no relaxation equation from Sec.~\ref{sec:spectrum}.
Source reconstruction under unknown group transformations is an established
statistical problem \cite{AbbePereiraSinger2018,BandeiraOrbit2023,Semerjian2025};
here its connection to the network's jointly preserved features is explicit.

A message $x\in\mathbb R^d$ originates at Alice, is relayed through the
network, and is later delivered to Bob. Bob has only the final observation,
with no intermediate measurements or independent copy of $x$.
Choose a known reference path from Alice to Bob, with translation $P$.
If the realized route has translation $T$, following that route and then
the reference path backward forms a closed walk based at Alice, with
return transformation $Q=P^{-1}T$. Thus $T=PQ$.
Undoing $P$ expresses Bob's observation in Alice's coordinates and leaves
the unknown factor $Q$. Alice is the \emph{root} of this closed-walk
description; no state of hers is held fixed. Bob need not lie on each
fundamental loop. Reciprocity permits inversion of a known path, but does
not identify an unrecorded path.

Assume that $Q$ is orthogonal, preserving vector lengths, and that its
distribution is known. In the source coordinates Bob observes
\begin{equation}
x\sim\mathcal N(0,s^2I_d),\qquad
y=Qx+\epsilon,\qquad
\epsilon\sim\mathcal N(0,\sigma^2I_d).
\label{eq:route_channel}
\end{equation}
The source and noise are independent centered Gaussian vectors with
variances $s^2>0$ and $\sigma^2\geq0$ per component, and neither has a
preferred direction. They are independent of the route. Noise is added
once at final measurement; undoing the orthogonal reference translation
preserves its distribution. The dictionaries remain specified inputs.

A \emph{decoder} is a rule $g(y)$ estimating the original source vector,
including its orientation and signed components. Its normalized error is
$\mathcal E(g)=\mathbb E\|x-g(y)\|^2/(ds^2)$, averaged over source,
route, and noise. Perfect reconstruction gives zero; always reporting
the zero vector gives error one. Write $c=s^2/(s^2+\sigma^2)$.
For any source-independent route distribution, let $M=\mathbb E[Q]$.
The smallest error over all decoders, including nonlinear ones, is attained by
\begin{equation}
\widehat x=cM^{\mathsf T}y,\qquad
\mathcal E=1-\frac{c}{d}\|M\|_F^2,
\label{eq:route_general}
\end{equation}
where $\|M\|_F^2$ is the sum of squared matrix entries.
Unequal route probabilities and dependent choices along a route are allowed.
This follows from the conditional-mean calculation below; SI \SIref{sec:s_route}
extends it to partial route records and rotationally invariant non-Gaussian sources.

\begin{figure*}[t!]
\includegraphics[width=\textwidth]{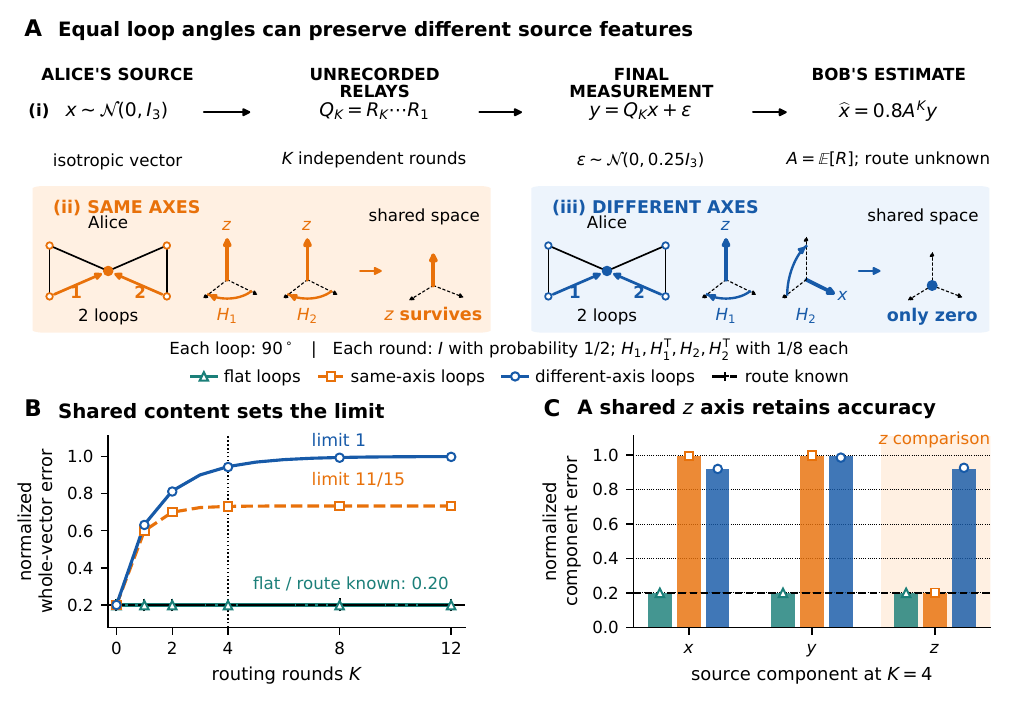}
\caption{\textbf{Identical loop spectra can give different recoverable features.}
\textbf{A(i)}~A source vector originates at Alice, undergoes $K$ unrecorded
loop choices, and is delivered to Bob for one noisy measurement. The known
final-path translation has been undone, so the displayed observation is
$y=Q_Kx+\epsilon$ in source coordinates. Bob has no original copy or
intermediate observations.
\textbf{A(ii)--(iii)}~Individuals carry three-component vectors; the two
triangles meet at Alice. Black edges carry $I_3$; colored edges 1 and 2
carry $H_1$ and $H_2$ along their arrows and inverse maps in reverse.
From Alice, following black edges and returning along colored edge $j$
applies $H_j$, shown in the matching rotation glyph.
\textbf{A(ii)}~Both maps are $R_z(\pi/2)$ and preserve the same $z$ axis.
\textbf{A(iii)}~The maps are $R_z(\pi/2)$ and $R_x(\pi/2)$: they preserve
perpendicular axes and have no nonzero jointly fixed vector. Both loops have spectrum
$\{1,i,-i\}$ in both conditions. Solid colored axes are preserved
directions; dashed black axes are coordinate guides, and curved arrows
indicate quarter-turns. Each round chooses $I$ with probability $1/2$
or either loop in either direction with probability $1/8$ each.
\textbf{B}~Smallest normalized whole-vector error,
Eq.~\eqref{eq:route_risk}, for $d=3$, $s^2=1$, $\sigma^2=1/4$.
Flat dictionaries and a recorded route give $1/5$; same-axis and
different-axis limits are $11/15$ and $1$. The vertical line selects
$K=4$ for \textbf{C}, whose component errors are normalized by $s^2$.
The highlighted $z$ comparison shows protection for aligned loops.
Lines and bars are exact predictions. Points average $120{,}000$
independent simulated source messages per condition; error bars are
pointwise $95\%$ Monte Carlo intervals and may be smaller than the symbols.
Recorded-route trials use the perpendicular-axis network. Source vectors
are withheld from the decoder and used only to measure error. SI \SIref{sec:s_route}
specifies the calculations.}
\label{fig:route}
\end{figure*}

To connect this limit explicitly to common ground, consider independent
routing rounds that sample every fundamental loop. Choose the
$b_1=E-N+1\geq1$ fundamental loops associated with a spanning tree,
based at Alice, with orthogonal
return matrices $H_1,\ldots,H_{b_1}$. Each round chooses the identity
with probability $1/2$, or one loop in either direction with probability
$1/(4b_1)$ each. A concrete realization makes these optional closed
excursions before delivery along the reference path. Each excursion acts
on the message currently carried; returning to Alice does not reset it
from the original copy. The identity can mean waiting or retracing an edge.

If $R_k$ is the choice in round $k$, then $Q_K=R_K\cdots R_1$ is the
accumulated transformation before delivery. Here $K$ counts routing
choices, not individual edges, elapsed time, or a configuration of the
root. At $K=0$ only the known reference path remains. The average for
one round is
\begin{equation}
A=\frac12 I_d+\frac{1}{4b_1}\sum_{j=1}^{b_1}(H_j+H_j^{\mathsf T}).
\label{eq:route_mean}
\end{equation}
Equal forward and reverse weights make $A$ symmetric; the identity weight
places its eigenvalues in $[0,1]$. This gives a simple exactly solvable
benchmark in which no fundamental loop is favored. Independence gives
$\mathbb E[Q_K]=A^K$, even when the loop matrices do not commute.
An individual message undergoes one realized $Q_K$; $A^K$ averages over
the histories Bob cannot distinguish. Such path averages are familiar
connection-graph objects \cite{SingerWu2012,Cloninger2024}.

\begin{theorem}[Reconstruction without a route record]
\label{thm:route_reconstruction}
For this routing protocol and Eq.~\eqref{eq:route_channel}, the smallest
error and an estimator attaining it are
\begin{equation}
\widehat x_K=cA^K y,\qquad
\mathcal E_K=1-\frac{c}{d}\tr(A^{2K}).
\label{eq:route_risk}
\end{equation}
Providing the route gives $\widehat x_{\rm rec}=cQ_K^{\mathsf T}y$
and error $\mathcal E_{\rm rec}=1-c$, for every connection.
If $r=d_{\rm share}$ is the joint fixed-space dimension, then
\begin{equation}
\frac{1-\mathcal E_K}{1-\mathcal E_{\rm rec}}
=\frac{\tr(A^{2K})}{d}
\longrightarrow\frac{r}{d}.
\label{eq:route_fraction}
\end{equation}
\end{theorem}

Equation~\eqref{eq:route_fraction} compares the reduction in error achieved
with and without a route record. After many unrecorded routing choices,
the fraction that remains is exactly the fraction of feature dimensions
preserved jointly by all loops. A known accumulated transformation
can be undone even when the loops are nonidentity.

Every route gives the same distribution of $y$ under
Eq.~\eqref{eq:route_channel}, so the observation does not reveal its route.
For a known $Q$, the Gaussian conditioning formula gives the conditional
mean of $x$ as $cQ^{\mathsf T}y$ \cite{Bishop2006}.
Averaging these means gives $cM^{\mathsf T}y$; a conditional mean
minimizes squared error over all decoders \cite{Bishop2006,Semerjian2025}.
For the independent rounds, $M=A^K$. The geometric connection follows from
\begin{equation}
v^{\mathsf T}(I_d-A)v
=\frac{1}{4b_1}\sum_j\|v-H_jv\|^2.
\label{eq:route_fixed}
\end{equation}
The right-hand side is the total squared change of a feature vector $v$
under the fundamental loops. It vanishes precisely when every loop
preserves $v$. The eigenvalue-one space of $A$ is therefore the joint
fixed space, and its other eigenvalues lie in $[0,1)$. Consequently,
$A^K$ approaches the projector onto that space.
The same limiting fraction holds for nonuniform symmetric loop weights,
provided every fundamental loop and the identity have positive probability.
Those weights change the approach to the limit. Supplemental Material,
Sec.~\SIref{sec:s_route}, gives the full proof of Theorem~\ref{thm:route_reconstruction}
and this nonuniform extension.

Fig.~\ref{fig:route}\textbf{A(i)} makes the source, unrecorded history,
and receiver distinct. In Fig.~\ref{fig:route}\textbf{A(ii)}, both loops preserve $z$;
in Fig.~\ref{fig:route}\textbf{A(iii)}, their axes are perpendicular. The graph, routing
probabilities, source, and noise are held fixed. Their errors approach
$1-c/3$ and $1$, while recording the route gives $1-c$ in both
(Fig.~\ref{fig:route}\textbf{B}). Thus equal individual-loop spectra
do not determine recovery without a route record.

Protection is selective (Fig.~\ref{fig:route}\textbf{C}). The $z$
component retains its route-recorded accuracy for aligned loops and loses
that protection for perpendicular loops. The $y$-component errors are
identical. If a nonzero matrix $B$ selects or combines the requested
components, the smallest error for reconstructing $Bx$, normalized by its
mean squared magnitude $s^2\|B\|_F^2$, is
$1-c\|BA^K\|_F^2/\|B\|_F^2$. Which content is requested matters even
when the total shared dimension is fixed.

These limits concern the specified source vector. With noiseless readout,
its length $\|x\|$ remains exactly recoverable, including when the smallest
oriented-vector error is one. Recovering such invariant properties is a
different task \cite{BandeiraOrbit2023,Semerjian2025}.
SI \SIref{sec:s_route} extends the geometric fraction to rotationally invariant
non-Gaussian sources and noise, including a spherical source with a
nonlinear decoder (Fig.~\SIref{fig:s9}); Fig.~\SIref{fig:s10} treats dictionary uncertainty.
Directionally structured sources, source-dependent routing, repeated
observations, additional reference cues, and lossy or nonlinear dictionaries
require further analysis. Changes of private coordinates leave the stated
predictions unchanged. The proposed applications are conditional uses of
the model.

\section{How the loop could be measured}
\label{sec:measure}

The loop geometry suggests what to measure, but measuring it requires more
than aligning private coordinates. Neural recordings from different animals
can be aligned to reveal shared patterns of activity \cite{Safaie2023};
such alignment does not by itself identify the translations used by an
interacting collective. An application would need pairwise dictionaries
validated on held-out observations, consistent local coordinates across
pairs, and uncertainty estimates. Here we specify summaries of their loop
action and bounds on what dictionary errors permit one to conclude.

\begin{proposition}[Loop fingerprint]
\label{prop:fingerprint}
For a chosen set $\mathcal R$ of measured feature sectors, we call the
ordered list
\[
\mathbf W_{\mathcal R}(C)=
\bigl(\chi_r(H_C)/d_r\bigr)_{r\in\mathcal R}
\]
the \emph{loop fingerprint}. Each entry is the trace of the loop matrix
in one sector, divided by that sector's dimension. Every entry is unchanged
by private relabeling. Two different lists therefore cannot describe
coordinate relabelings of the same loop transformation.
\end{proposition}

Private relabeling conjugates the loop matrix, leaving its trace unchanged
(Proposition~\ref{prop:conjugacy}); Supplemental Material, Sec.~\SIref{sec:s9},
gives the argument.

\begin{figure*}[t]
\includegraphics[width=\textwidth]{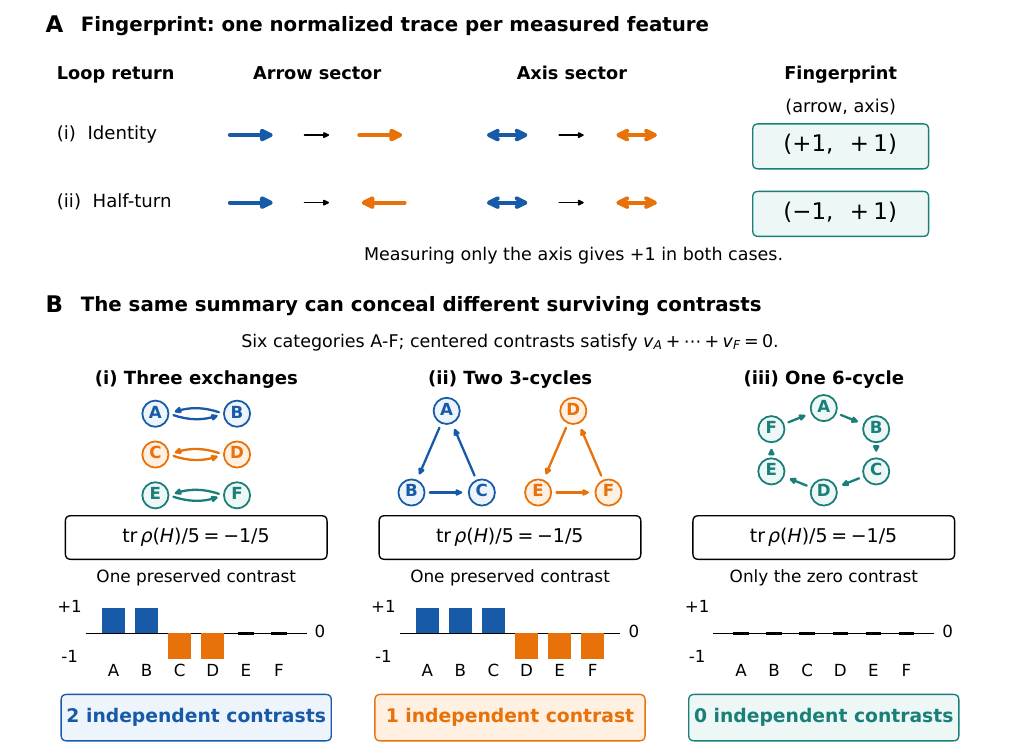}
\caption{\textbf{What a loop fingerprint reveals, and what it misses.}
\textbf{A}~The fingerprint is an ordered list of normalized traces, here
(arrow, axis). Blue pictograms show the initial feature and orange its
loop return; black arrows indicate the transformation.
\textbf{A(i)}~Identity gives $(1,1)$.
\textbf{A(ii)}~A half-turn gives $(-1,1)$.
Keeping only the axis entry makes these returns indistinguishable.
\textbf{B}~Six category amplitudes form a centered contrast when their sum
is zero. Directed cycles show which labels are permuted. A preserved
contrast has equal amplitudes throughout each permutation cycle.
\textbf{B(i)}~Three exchanges $(AB)(CD)(EF)$ preserve two independent
centered contrasts; bars show $(1,1,-1,-1,0,0)$.
\textbf{B(ii)}~Two cycles $(ABC)(DEF)$ preserve one, shown as
$(1,1,1,-1,-1,-1)$.
\textbf{B(iii)}~A six-cycle $(ABCDEF)$ forces all amplitudes equal,
leaving only the zero centered contrast. All three permutations move every
label, so their trace in the five-dimensional centered sector is $-1$ and
their normalized trace is $-1/5$. Positive bars are blue, negative bars
orange, and black marks zero. Equal fingerprints need not mean equal
preserved content.}
\label{fig:fingerprintguide}
\end{figure*}

The fingerprint records one number per measured sector, rather than the
whole return matrix. For the food-direction example, order its entries as
(arrow, axis). The identity return gives $(1,1)$
(Fig.~\ref{fig:fingerprintguide}\textbf{A(i)}), whereas a half-turn gives
$(-1,1)$ (Fig.~\ref{fig:fingerprintguide}\textbf{A(ii)}): it reverses both arrow components and preserves
both axis-sector components. Measuring both sectors distinguishes these
returns. Measuring only the axis keeps just the second entry, $1$, and
cannot distinguish them. Choosing what to measure therefore determines
which loop differences the fingerprint can reveal.

Equal fingerprints mean that the chosen measurements agree, not that the
loops act identically. The trace adds all eigenvalues; it does not in
general tell us how many equal $1$, which is what counts preserved
directions. Figure~\ref{fig:fingerprintguide}\textbf{B(i)--(iii)} shows
three six-category permutations. Every label moves, so each full
permutation matrix has trace zero. Removing the all-equal sector, whose
trace is one, leaves trace $-1$ on five centered coordinates: all three
fingerprints are $-1/5$. Yet the permutations group categories into
three, two, and one cycles. A preserved contrast is constant within each
cycle, leaving two, one, and zero independent contrasts after imposing
zero sum. Their identical fingerprints conceal different fixed spaces.

Additional measured sectors can resolve ambiguities; Supplemental Material,
Sec.~\SIref{sec:s9} and Fig.~\SIref{fig:s8}, give a complete three-category example and the
established character criterion \cite{Serre1977,BrockerDieck1985}.
For several network loops, a further issue arises: even their individual
spectra omit the relative orientation of the features they preserve.
The aligned and different-axis loops in Fig.~\ref{fig:route}\textbf{A(ii)--(iii)}
illustrate why sharing depends on their \emph{joint} fixed space.
Composite loops retain joint geometric information \cite{Gao2021};
additional measured sectors can distinguish actions left unresolved by
one sector \cite{GaoZhao2019}.

\begin{proposition}[Dictionary-to-loop error bounds]
\label{prop:certificate}
Assume true and estimated dictionaries, $U_{ij}$ and $\widehat U_{ij}$,
are unitary and reciprocal on the same graph with the same weights.
If their error in sector $r$ obeys
$\lVert\rho_r(U_{ij})-\rho_r(\widehat U_{ij})\rVert_2
\leq\varepsilon_{ij}=\varepsilon_{ji}$ on every edge, then
\[
\lVert\rho_r(H_C)-\rho_r(\widehat H_C)\rVert_2
\leq\sum_{e\in C}\varepsilon_e .
\]
Errors are compared in the same private coordinates; each loop traversal
contributes to the sum, including repeated uses of an edge.
Here $\|\cdot\|_2$ is the operator norm: the largest norm of the image of a unit vector.
Every ordered eigenvalue of $L_U^r$ changes by at most
$\delta=\max_i\sum_jw_{ij}\varepsilon_{ij}$.
\end{proposition}

The bounds follow from expanding the difference of matrix products and
from the Hermitian eigenvalue perturbation bound \cite{HornJohnson2013};
Supplemental Material, Sec.~\SIref{sec:s9}, gives both proofs and numerical examples.

Loop uncertainty thus accumulates no faster than the sum of the dictionary
uncertainties along the route. An estimated loop farther from the identity
than that sum must be genuinely nonidentity. Likewise, an estimated
Laplacian eigenvalue larger than $\delta$ has a positive true eigenvalue
at the same ordered index. This constrains the shared dimension without
identifying individual eigenvectors. A near-zero estimate alone does not establish
an exact shared mode.

\section{Discussion}
\label{sec:discussion}

Much of collective-behavior theory has asked how individuals reach
agreement while taking a common frame for granted
\cite{Vicsek1995,DeGroot1974,HegselmannKrause2002}. This leaves a prior
question implicit: when individuals carry private representations, which
collective states can exist at all? We make this question explicit and
quantitative. For reciprocal unitary translations, understanding within
every pair need not produce compatibility across the group: the features
preserved jointly by its loops determine its common ground. Under the
dynamics and reconstruction protocol studied here, this relational geometry
also governs how incompatible patterns decay and what content remains
recoverable when a message's route is unknown. Shared-frame comparison
becomes the globally consistent case within a broader theory of what a
collective can share, retain, and recover.

Our work can impact several disciplines. For network science, a link
carries a third attribute: besides who interacts and how strongly, what
happens to represented content as it crosses. Two collectives with identical
individual state spaces, graphs, and weights can support different shared
states because their translations compose differently. The distinction
resides in their loop transformations considered jointly, up to a common
conjugation---the holonomy description of gauge connections on a graph
\cite{Gao2021}. For collective decision-making, failure acquires a mechanism
that requires neither noise nor bias: the nonzero shared state the members
are asked to reach may not exist, and Eq.~\eqref{eq:kernel} identifies exactly
which contrasts survive. For the study of behavior,
Theorem~\ref{thm:functional} makes ``the world as represented'' physically
consequential: the same loop mismatch can obstruct sharing in one
representation and be invisible in another, so what a collective represents
determines which relational differences it detects. The coarsening relation
$K_F\subseteq K_Q$ shows that discarding distinctions can make additional
mismatches invisible. It thereby opens an evolutionary question: when should
selection repair a dictionary, rely on it less, or stop representing the
failing distinction? These adaptive possibilities remain to be investigated.

\emph{Two meanings of reciprocity.} Empirical work on interaction laws in
moving animal groups infers effective forces from responses to neighbors
\cite{Katz2011}. Whether the \emph{force law} is reciprocal is a question about
symmetry of influence under exchanging
the interacting agents. Such symmetry alone does not establish a
conservative force field. That is logically independent of
reciprocity of the \emph{dictionary}, Eq.~\eqref{eq:reciprocity}: a perfectly
Hamiltonian pair interaction is compatible with a non-reciprocal learned
translation, and a
non-potential force law is compatible with perfectly reciprocal dictionaries.
Dictionary reciprocity is an assumption of
Theorem~\ref{thm:kernel}; where it fails, the directed operator of
Remark~\ref{rem:directed} applies and no equivalence is claimed. Neither
meaning of reciprocity should be used as evidence about the other.

\emph{Relation to prior mathematics.} Connection Laplacians and their
kernels \cite{SingerWu2012,Bandeira2013}, synchronization through holonomy
\cite{Gao2021}, angular synchronization \cite{Singer2011}, and Wilson-loop
invariants \cite{Wilson1974,Kogut1979,Wegner1971} provide the geometric
foundations. Cellular-sheaf consensus describes private opinion spaces,
communication maps, and diffusion toward compatible states
\cite{HansenGhrist2021}. Matrix-weighted consensus studies shared and
clustered states \cite{Trinh2018}, though its positive-semidefinite
matrix weights multiplying $(a_i-a_j)$ generally differ from the
unitary transports in $(a_i-U_{ij}a_j)$. Frustrated $XY$ models provide
an established physical setting in which bond phases make loop mismatch
consequential \cite{Shastry1982}. Gauge fields and cycle holonomy also enter
collective oscillator dynamics \cite{BeuriaChembrolu2026,TorresHugas2026},
and loop geometry is studied in learned representations
\cite{Sevetlidis2026,GroverBourgerie2026,Javidnia2026}.

The contribution developed here connects these foundations to a feature-resolved
physical question: which features can be shared through a given system of reciprocal
translations between private representations? The representation-dependent
criterion, cross-representation examples, and finite-time predictions give
that question a common quantitative formulation. Reconstruction is an
operational consequence of the same geometry. Building on inference under
group actions \cite{AbbePereiraSinger2018,Semerjian2025},
Theorem~\ref{thm:route_reconstruction} connects the network's shared feature
space to an exact source-vector error under the stated observation model.
The comparison in Fig.~\ref{fig:route} isolates the role of joint loop
action at fixed individual-loop spectra, source statistics, and noise.
Providing the route restores the noise-limited error. Thus a structural
property of the relations determines a limit on recovering represented
content.

\emph{Scope and next questions.} The connection-Laplacian equivalences assume
connected graphs, symmetric positive weights, and reciprocal unitary
translations on matched sectors. Nonreciprocal, nonisometric, singular,
or unequal-dimensional maps still define the difference operator in
Remark~\ref{rem:directed}; their spectral and reconstruction properties
require separate analysis. Time-varying dictionaries also lie outside
the fixed-operator dynamics studied here.
The ring spectrum is a worked example on one topology, and
the counting law uses the stated uniform finite-group measure. Dynamical
claims are limited to the specified linear relaxation and message-routing
protocols. The gauge correspondence concerns classical operators; it
transfers mathematical structures without attributing quantum behavior or
quantum cognition to organisms. The maps are specified inputs. Deriving how learning shapes
their distribution, how changing relations alter collective organization,
and when selection favors repair or coarser representations are further
theoretical questions. Empirical identification of
holonomy in a living collective remains open; the present uncertainty
bounds specify what assumed dictionary errors permit one to conclude.

{\widowpenalty=10000
\emph{Conclusion.} Newton's frame served physics well for planets, and it has served collective
behavior as scaffolding. But a collective of private worlds is a system of
frames, and the physics that survives the removal of the scaffold is the
physics of how those frames compose. What a collective can hold in common
is written in its loops, within the collective, not in an observer's frame.
Such a frame, however, remains a useful description whenever the collective
lies in the flat sector. How often living collectives reach that flat sector
through learning, evolution, or other processes remains a question for
future work.\par}

\begin{acknowledgments}
The author acknowledges funding from Deutsche Forschungsgemeinschaft
(German Research Foundation) under Germany's Excellence Strategy---
EXC 2117--422037984.

OpenAI Codex (ChatGPT 5.6 and 6) and Anthropic Claude (Fable 5) assisted
with mathematical analysis, programming and numerical checks, figure
preparation through plotting-code development, writing, and manuscript
review and editing. No AI-generated image assets were used. The author
directed the work, reviewed and verified all AI-generated material, and
takes responsibility for the manuscript's content.
\end{acknowledgments}

\section*{Data and code availability}
The code and numerical data supporting this work are provided in the
accompanying Supplemental Material \cite{SupplementalMaterial}. The code
includes the model parameters and reproduces all main-text and
supplementary figures.
\nocite{Diestel2017,Hatcher2002,HornJohnson2013,DummitFoote2004,BrockerDieck1985,Gao2021,Wilson1974,Kogut1979}
\nocite{SingerWu2012,Bandeira2013,Serre1977,HansenGhrist2021,MardiaJupp1999,Bishop2006,NISTDLMF,OlfatiSaberMurray2004}
\nocite{Altafini2013,Harary1953,DeGroot1974,AharonovBohm1959,Hofstadter1976,Higham2008,Tsitsulin2018,Teschl2009}
\nocite{Sikora2012,AbbePereiraSinger2018,BandeiraOrbit2023,Cloninger2024,Semerjian2025,Owen2013,JohnsonKotzBalakrishnan1995,Mezzadri2007}
\let\OriginalBibliographyFont\bibfont
\renewcommand{\bibfont}{\OriginalBibliographyFont\fontsize{9}{10}\selectfont}
\bibliography{refs}

\end{document}